\documentclass[preprint,aps,showpacs,floatfix]{revtex4}
\usepackage{amsmath,amssymb,graphicx,epsfig}

\newcommand{\vv}{{\bf {v}}}
\newcommand{\rr}{{\bf {r}}}
\newcommand{\pp}{{\bf {p}}}

\newcommand{\ephi}{{\mbox{\boldmath $\hat e_{\phi}$}}}
\newcommand{\ep}{{\mbox{$\epsilon_p$}}}
\newcommand{\eF}{{\mbox{$\epsilon_F$}}}
\newcommand{\vF}{{\mbox{$v_F$}}}
\newcommand{\pF}{{\mbox{$p_F$}}}

\numberwithin{equation}{section}

\begin{document}

\title{\bf Ballistic propagation of thermal excitations
near a vortex in superfluid $^3$He-B}

\author{C. F. Barenghi${}^1$, Y. A. Sergeev${}^2$ and N. Suramlishvili${}^{1,3}$ }

\affiliation {
${}^1$School of Mathematics and Statistics,
Newcastle University, Newcastle upon Tyne, NE1 7RU
${}^2$School of Mechanical and Systems Engineering,
Newcastle University, Newcastle upon Tyne, NE1 7RU
${}^3$Andronikashvili Institute of Physics, Tbilisi, 
0177, Georgia}
\date {\today}

\begin {abstract}
Andreev scattering of thermal excitations is a powerful tool for studying
quantized vortices and turbulence in superfluid $^3$He-B at very low
temperatures. We write Hamilton's equations for a quasiparticle in the 
presence of a vortex line, determine
its trajectory, and find under wich conditions it is Andreev reflected. 
To make contact with experiments, we generalize our results to the 
Onsager vortex gas, and find values of the intervortex spacing 
in agreement with less rigorous estimates.
\end{abstract}

\maketitle

\section{Motivation}

\renewcommand{\theequation}{\arabic{section}.\arabic{equation}}

Superfluid turbulence consists of a disordered tangle of 
quantized vortex filaments which move under the velocity field of
each other\cite{Donnelly-book,BDV-book}. If the temperature, $T$ is
sufficiently smaller than the critical temperature, $T_c$, then the
normal fluid can be neglected and the vortices do not experience any
friction effect\cite{BDV}. The simplicity of the vortex structures
(discrete vortex lines) and the absence of dissipation mechanisms,
such as friction and viscosity, make superfluid turbulence a
remarkable fluid system, particularly when compared to turbulence in
ordinary fluids. Current experimental, theoretical and numerical
investigations attempt to determine the similarities and the
dissimilarities between superfluid turbulence and ordinary
turbulence. Questions which are currently addressed concern (i) the
existence of a Kolmogorov energy cascade at length scales larger
than the typical intervortex spacing\cite{Vinen-Niemela,Hulton},
(ii) the existence of a Kelvin wave cascade at length scales smaller
than the Kolmogorov
length\cite{Kivotides-cascade,Vinen-cascade,Kozik-cascade,Nazarenko-cascade}
followed by (iii) acoustic emission at even shorter length
scales\cite{Vinen-sound,Parker}, (iv) the possible existence of a
bottleneck\cite{Nazarenko-bottleneck,Svistunov-bottleneck} between
the Kolmogorov cascade and the Kelvin wave cascade, (v) the nature
of the fluctuations of the observed vortex line
density\cite{Roche,Roche-Barenghi,Tabeling,Bradley3} and (vi) their
decay\cite{towed-grid,decay}, (vii) whether there are two forms of
turbulence\cite{Vinen-Kolmogorov}, a structured one, which consists
of many length scales (Kolmogorov turbulence), and an unstructured,
more random one (Vinen turbulence), (viii) the effects of rotation
on turbulence\cite{Finne,Mongiovi,Tsubota,Eltsov}. Most of these
questions refer to the important limit $T/T_c \ll 1$, where fundamental
distinctions between  a perfect Euler fluid and a superfluid becomes
apparent\cite{Barenghi}.

Superfluid turbulence experiments are currently performed in both
$^4$He \cite{Golov,Roche-Barenghi,McClintock,Skrbek,Roche} and in $^3$He-B
\cite{Bradley3,Fisher,Finne,Yano}. In the last few years it has been
recognized that, to make progress in answering the above questions,
it is necessary to develop better measurement techniques which are
suitable for turbulence in quantum fluids. In $^4$He, the
application of the classical PIV
method\cite{VanSciver-piv,Bewley-piv,Kivotides-piv} was a
breakthrough.  In $^3$He-B, a non--classical, powerful measurement
technique which is suitable in the limit $T/T_c \ll 1$ is
the Andreev scattering\cite{Fisher}, developed at the University of
Lancaster.

This article is concerned with the Andreev scattering. The plan of
the paper is the following. In Section II we shall describe the
basic ideas behind the Andreev scattering and review what is a
quantized vortex line. In Section III we shall write down the
governing equations of motion. In Section IV we shall determine the
ballistic trajectories of excitations in the vicinity of the
velocity field of a vortex line, and, in Section V, we shall study
the transport of heat by ballistic quasiparticles through a tangle
of vortices. Section VI will apply our result to the
current experiments. Finally, in Section VII, we shall draw the
conclusions.

\section{Andreev scattering and quantized vortices}

\renewcommand{\theequation}{\arabic{section}.\arabic{equation}}

The study of the motion of quasiparticle excitations in a superfluid
was pioneered by Andreev\cite{Andreev}. Consider an excitation which
moves in the direction of increasing excitation gap.
The excitation propagates at constant energy, and gradually
reaches the minimum of the rising excitation spectrum,
where its group velocity becomes zero. Thereafter it retraces its path
but as an excitation on the other side of the minimum.
An incoming quasiparticle is thus reflected as a
quasihole and an incoming quasihole is reflected as a quasiparticle.
The effect is a consequence of the fact that the minimum of the energy spectrum
of the excitation lies at nonzero momentum.

The case of $p$-wave triplet pairing appropriate to superfluid
$^3$He has been discussed by various authors studying the
interaction of excitations with the boundaries\cite{KR}, motion of
quasiparticles through the $A$-$B$ phase boundary in
$^3$He\cite{Yip}, ballistic motion of quasi--particle in slow
varying textural field of $^3$He-A\cite{Leggett}, scattering
of ballistic quasiparticles in $^3$He-B by a moving solid surface
\cite{Guenault,Enrico}, and calculation of the friction force
on quantized vortices\cite{Kopnin-Kravtsov,Stone}. Ref. \cite{Stone}
and \cite{Volovik} are concerned with Andreev reflection within the 
vortex core and therefore apply to the bound states. Our concern is the
propagation of thermal excitations outside vortex cores.

Collisions between the quasiparticles can cause some spreading of
the incoming beam. However, the spreading can be made arbitrary
small by lowering the density of the excitations, that is to say, by
lowering the temperature. At low enough temperatures the mean free
path exceeds the dimensions of the experimental cell and we can
consider undamped excitations moving along straight paths until they
hit a boundary or any potential barrier, particularly a barrier
formed by a vortex. Andreev reflection of excitations thus gives the
opportunity to probe flows in superfluid $^3$He at ultra--low
temperatures. The most fruitful and promising application of Andreev
scattering is thus superfluid turbulence in $^3$He-B in the low
temperature limit, that is to say for $T/T_c \leq 0.4~\textrm K$
\cite{Bradley3,Fisher,Bradley1,Bradley2,QFS-Fisher}.

Superfluid $^3$He-B is described by a macroscopic wave function,
called the order parameter, with a well defined phase $\phi$. The
superfluid velocity ${\bf v}_s$ is proportional to the gradient of
the phase,
\begin{equation}
{\bf v}_s=\frac{\hbar}{2m}\nabla\phi,
\end{equation}
where $m$ is the mass of one $^3$He atom. Consequently, in contrast
to classical fluids, superfluid motion is irrotational and vorticity
exist only in the form of quantized vortices.
Quantized vortices are line defects around which the phase $\phi$
changes by $2\pi$. The superfluid order parameter is distorted
within the relatively narrow core of the vortex, and the superfluid
flows around the core with speed which is inversally proportional to
the distance from the vortex core. Since both the real and the
imaginary parts of the order parameter are zero on the axis of a
vortex, vortex lines can be considered as topological defects.
Vortices cannot terminate in the middle of the flow, so they are
either closed loops or extend to the walls.

Superfluid turbulence consists of a tangle of quantized  vortices. The
complex flow field within the tangle acts as a  potential barrier for
quasiparticles, causing the Andreev reflection
of a fraction of a beam of thermal excitations incident upon the
tangle. The use of Andreev scattering as a visualization technique
of ultra--low temperature turbulence requires to find out exactly
what happens to a single quasiparticle which moves in the
velocity field of a vortex, which is what we set out to do.

\section{Equations of motion of thermal excitations}

\renewcommand{\theequation}{\arabic{section}.\arabic{equation}}

Our first aim is to formulate, in the  $(x,\,y)$-plane, the
equations of motion of a single excitation moving in the velocity
field of a single straight vortex which we assume to be fixed and
aligned along the $z$-axis. We are thus concerned with a
two-dimensional problem only. The quantities (here and below the
numerical values of the quantities are taken at the $0$ bar
pressure\cite{Greywall}) which are necessary to describe the motion of the
excitation are the Fermi velocity, $\vF \approx 5.48 \times 10^3~\rm
cm/s$, the Fermi momentum, $\pF=m^* \vF\approx 8.28 \times
10^{-20}~\rm g~ cm/s$, and the Fermi energy,
$\eF=\pF^2/(2m^*)\approx 2.27 \times 10^{-16}~ \rm erg$. The
quantity
\begin{equation}
\ep=\frac{p^2}{2m^*}-\eF \label{eq:ep}
\end{equation}
is the "kinetic" energy of the excitation measured with respect to
the Fermi energy, $\eF$, where $m^*\approx 3.01\times m=1.51 \times
10^{-23}~\rm g$ is the effective mass of the excitation, and $\pp$
the momentum, $p=\vert \pp \vert$. Let $\Delta_0$ be the magnitude
of the superfluid energy gap. Near the vortex axis, at radial
distances $r$ smaller than the zero--temperature coherence length
$\xi_0 = \hbar v_F/\pi\Delta_0\approx0.8 \times 10^{-5}~\rm
cm$, the energy gap falls to zero and can be
approximated by $\Delta(r) \approx \Delta_0\tanh(r/\xi_0)$
\cite{Bardeen,TsunetoBook}. Since we are mainly concerned with what happens for
$r \gg \xi_0$, we neglect the spatial dependence of the energy gap
and assume the constant value $\Delta_0=1.76k_B T_c \approx 2.43
\times 10^{-19}~\rm erg$.

Using polar coordinates $(r,\,\phi)$ in the $(x,\,y)$ plane , the velocity field 
of a superfluid vortex set along the $z-$axis is
\begin{equation}
\vv_s=\frac{\kappa}{2 \pi r} \ephi,
\label{eq:vs}
\end{equation}
where
\begin{equation}
\kappa=\frac{h}{2m}=\frac{\pi\hbar}{m}=0.662\times
10^{-3}~\textrm{cm}^2\textrm{/s} \label{eq:kappa}
\end{equation}
is the quantum of circulation, and $\ephi$ is the unit vector in the
azimuthal direction on the $(x,\,y)$-plane.

In the presence of the vortex, the energy of the excitation becomes
\begin{equation}
E=\sqrt{\ep^2+\Delta_0^2}+\pp \cdot \vv_s. \label{eq:E}
\end{equation}
In writing Eq.~(\ref{eq:E}), the spatial variation of the order
parameter is not taken into account for the sake of simplicity. We
also assume that the interaction term $\pp \cdot \vv_s$ varies on a
spatial scale which is larger than $\xi_0$, and that the excitation
can be considered a compact object of momentum $\pp=\pp(t)$,
position $\rr=\rr(t)$, and energy $E=E(\pp,\,\rr)$. This gives us
the opportunity to use the method developed in Ref.\cite{Leggett},
and consider Eq.~(\ref{eq:E}) as an effective Hamiltonian, for which
the equations of motion are
\begin{equation}
\frac{d\rr}{dt}=\frac{\partial E(\pp,\rr)}{\partial \pp}
=\frac{\ep}{\sqrt{\ep^2+\Delta_0^2}} \frac{\pp}{m^*} + \vv_s,
\label{eq:hamilton1}
\end{equation}
\begin{equation}
\frac{d\pp}{dt}=-\frac{\partial E(\pp,\rr)}{\partial \rr}
=-\frac{\partial}{\partial \rr}(\pp \cdot \vv_s).
\label{eq:hamilton2}
\end{equation}

Eqs.~(\ref{eq:hamilton1}) and (\ref{eq:hamilton2}) have one
immediate integral of motion, the energy:
\begin{equation}
E(\pp,\,\rr)=E=\textrm{constant}. \label{eq:integral}
\end{equation}
Eq.~(\ref{eq:hamilton1}) represents the group velocity of the
excitation in the velocity field of the vortex. Excitations such
that $\ep>0$ are called quasiparticles, and excitations such that
$\ep<0$ are called quasiholes. The right-hand-side of
Eq.~(\ref{eq:hamilton2}) is thus the force acting on the excitation.

\section{Propagation of excitation in the velocity field of a vortex}

\renewcommand{\theequation}{\arabic{section}.\arabic{equation}}

We want to determine the trajectory of an excitation which moves in
the two-dimensional velocity field of the vortex. It is convenient
to rewrite the Hamiltonian, Eq.~(\ref{eq:E}), and Hamilton's
equations~(\ref{eq:hamilton1}) and (\ref{eq:hamilton2}) in polar
cordinates $(r,\,\phi)$. We notice that the system consisting of a
single excitation and a single vortex has a second integral of
motion: the component of the angular momentum in the $z$-direction,
perpendicular to the plane of motion, $(x,\,y)$. Consequently, we
can introduce two pairs of canonically conjugated variables,
$(p_r;\,r)$ and $(J=p_{\phi}r;\,\phi)$, where $p_r$ and $p_{\phi}$
are the radial and azimuthal components of $\pp$, and $J$ is the
angular momentum. Since $J$ is constant, it is convenient to write
it in the form $J=\pF \rho_0$, thereby defining the constant
$\rho_0$ for a particular trajectory. Under special initial
conditions, as we shall see, $\rho_0$ can be interpreted as the
impact parameter.

Eqs.~(\ref{eq:E}), (\ref{eq:ep}) and (\ref{eq:hamilton1}) become
\begin{equation}
E=\sqrt{\ep^2 + \Delta_0^2}+\pF \rho_0\frac{\kappa}{2 \pi r^2}\,,
\label{eq:E-new}
\end{equation}
\begin{equation}
\ep=\frac{p_r^2}{2m^*}+ \frac{(\pF \rho_0)^2}{2 m^* r^2}-\eF,
\label{eq:ep-new}
\end{equation}
\begin{equation}
{\dot
r}=\frac{dr}{dt}=\frac{\ep}{\sqrt{\ep^2+\Delta_0^2}}\frac{p_r}{m^*}\,,
\label{eq:r-dot}
\end{equation}
\begin{equation}
{\dot \phi}=\frac{d\phi}{dt}=\frac{\ep}{\sqrt{\ep^2+\Delta_0^2}}
\frac{\pF \rho_0}{m^* r^2}+ \frac{\kappa}{2 \pi r^2}\,.
\label{eq:phi-dot}
\end{equation}

By setting $dE/dt=0$ and using Eq.~(\ref{eq:r-dot}) we find
\begin{equation}
\dot{\ep}=\frac{d\ep}{dt}=\pF \rho_0
\frac{p_r}{m^*}\frac{\kappa}{\pi r^3}\,, \label{eq:ep-dot}
\end{equation}
and from Eq.~(\ref{eq:ep-new}) we have
\begin{equation}
\vert p_r \vert =\pF
\left(1+\frac{\ep}{\eF}-\frac{\rho_0^2}{r^2}\right)^{1/2}\,.
\label{eq:pr}
\end{equation}

Eqs.~(\ref{eq:E-new})-(\ref{eq:pr}) form a closed set which allows
us to determine the trajectory of the excitation.

It is apparent from Eq.~(\ref{eq:r-dot}) that a quasiparticle
incident upon the vortex has $\ep>0$ and $p_r<0$, whereas
a quasiparticle moving away from the vortex has $\ep>0$ and
$p_r>0$. Vice--versa, a quasihole incident upon the vortex
is characterized by $\ep<0$ and $p_r>0$, whereas a quasihole moving
away from the vortex has $\ep<0$ and $p_r<0$.

\begin{figure}
\centering \epsfig{figure=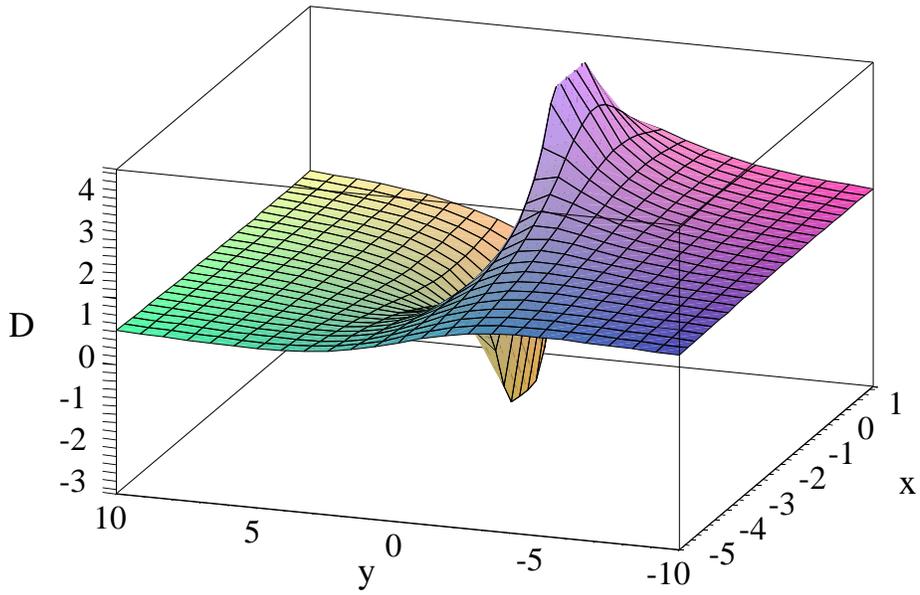,height=3.5in,angle=0}
\caption{(Color online) The dimensionless effective potential
$D=\Delta_{eff}(r/\xi_0)/\Delta_0$ seen by quasiparticles with
momentum parallel to the $x$-axis and moving from $x=-\infty$. 
The dimensionless coordinates $x$ and $y$ are in units of $\xi_0$.}
\label{fig:potential}
\end{figure}

Later we shall consider a quasiparticle which leaves a point of the
wall of the cylindrical experimental cell; this quasiparticle is
initially characterized by $r=R$ (where $R$ is the radius of the
cell), $p=\pF$ and $p_r<0$. The axis of the vortex will still be at
the centre of the coordinate system. In such case the quasiparticle
with initial momentum directed along the $x$-axis will feel the
effective pairing potential $\Delta_{eff}\approx\Delta_0-\pF
y\kappa/(2\pi r)$ (Fig.~\ref{fig:potential}).

It is obvious from Eq.~(\ref{eq:r-dot})
that unless $\rho_0$ is exactly zero ($J=0$), the radial
velocity of the excitation will eventually vanish. This may happen
either because $p_r=0$ (classical turning point) or because
$\ep=0$ (Andreev turning point).

It can be seen from Eqs.~(\ref{eq:E-new}) and (\ref{eq:pr}) that the
classical turning point is reached first when
\begin{equation}
E>\Delta_0 + \pF \frac{\kappa}{2 \pi \rho_0}\approx
\Delta_0\biggl(1+\frac{3\pi\xi_0}{2\rho_0}\biggr)
\label{eq:classical-turning-point}
\end{equation}
(here and in the equations below the numerical factor
3 is introduced by the ratio between the effective mass of quasiparticle
and the bare mass of a $^3$He atom: $m^*/m\approx 3$)
in which case a quasiparticle with this energy follows a trajectory
which is of the "normal" type: the quasiparticle retains its
"particle" nature and moves past the vortex, across the experimental
cell to the wall on the opposite side. On the contrary, a
quasiparticle with energy $E$ such that
\begin{equation}
\Delta_0<E<\Delta_0 + \pF\frac{\kappa}{2 \pi \rho_0}\approx
\Delta_0\biggl(1+\frac{3\pi\xi_0}{2\rho_0}\biggr)
\label{eq:Andreev-turning-point}
\end{equation}
reaches the Andreev turning point first, undergoes Andreev
reflection, and returns to a point near its starting point after
changing its nature and becoming a quasihole.

Of these two cases, our concern is the case of Andreev reflection.
We first determine the locus of Andreev turning points, defined
by the minimum radial distance from the vortex core:
\begin{equation}
r_{min}=\left( \frac{\kappa}{2 \pi}
\frac{\pF \rho_0 }{(E-\Delta_0)}\right)^{1/2}
=\left( \frac{3\pi\xi_0 \rho_0}{2}
\frac{\Delta_0}{(E-\Delta_0)}\right)^{1/2}.
\label{eq:rmin}
\end{equation}

Consider a quasiparticle which has reached
$r=r_{min}$. At this point
the radial velocity $\dot r$ vanishes, but the excitation
does not stop.  It has still a nonzero azimuthal velocity,
$r \dot \phi$. Thereafter the excitation propagates as a
quasihole (characterized by a negative value of $\ep$).

In order to calculate the trajectory of a reflected quasiparticle
it is convenient to simplify the governing equations of motion
using the fact that at the ultra--low temperatures which interest
us, $T \ll T_c$, most quasiparticles have energies
$\ep \ll \Delta_0$. We can then make the following
approximation:

\begin{equation}
\sqrt{\ep^2 + \Delta_0^2} 
\approx \Delta_0 + \frac{\ep^2}{2\Delta_0}
= \Delta_0 +\frac{(p^2-p_F^2)^2}{8 m^{*2} \Delta_0} 
\approx \Delta_0 +\frac{(p-p_F)^2}{2\Delta_0/v_F^2}\,. \label{eq:app}
\end{equation}

This spectrum is similar to Landau's spectrum of excitations in superfluid
He~II near the roton minimum ($p=p_0$), 
$E \approx \Delta_0+(p-p_0)^2/(2 m_r)$ 
(where $m_r$ is the effective roton mass), which was used to 
calculate the mutual friction force \cite{Samuels}; note that in
Eq.~(\ref{eq:app}) the role of the roton mass is played by the ratio
$\Delta_0/v_F^2$.

Using Eqs.~(\ref{eq:app}), (\ref{eq:r-dot}) and (\ref{eq:phi-dot}),
and the smallness of the ratios $\ep/\eF$ and $\Delta_0/\eF$, we
obtain
\begin{equation}
dt=\frac{m^* \Delta_0}{\ep p_r}dr= -\frac{m^*}{\pF}\frac{r_{min}}
{(3\pi\xi_0 \rho_0)^{1/2}} \frac{r^2 dr}{(r^2-r_{min}^2)^{1/2}
(r^2-\rho_0^2)^{1/2}}\,, \label{eq:dt}
\end{equation}
\begin{equation}
d\phi=-\left (
\frac{\rho_0} {r^2 (1-\rho_0^2/r^2)^{1/2} }
\pm
\frac{3 b r_{min}}
{2 (3\pi\xi_0 \rho_0)^{1/2}
(r^2-r_{min}^2)^{1/2} (r^2-\rho_0^2)^{1/2}} \right ) dr,
\label{eq:dphi}
\end{equation}
where $b=\hbar/\pF$, the sign plus is used for quasiparticles and
the sign minus for quasiholes.

From Eqs.~(\ref{eq:dt}) and (\ref{eq:dphi}) we obtain the Andreev
return time $\tau$ of the excitation (the time it takes to travel
from the radial distance $R$ to the Andreev reflection point and
back) and the Andreev reflection angle $\Delta \phi$:
\begin{equation}
\tau=2 \frac{r_{min}}{\vF} \frac{1}{(3\pi\xi_0 \rho_0)^{1/2}}
\int_{r_{min}}^R \frac{r^2 dr}{ (r^2-r_{min}^2)^{1/2}
(r^2-\rho_0^2)^{1/2}}\,, \label{eq:tau-1}
\end{equation}
\begin{equation}
\Delta \phi= 3 \frac{b r_{min}}{(3\pi\xi_0 \rho_0)^{1/2}}
\int_{r_{min}}^R \frac{dr}{ (r^2-r_{min}^2)^{1/2} (r^2
-\rho_0^2)^{1/2}}\,. \label{eq:deltaphi-1}
\end{equation}

The evaluation of these elliptic integrals is shown in the
Appendix. We obtain
\begin{equation}
\tau=2 \frac{r_{min}}{\vF} \frac{1}{(3\pi\xi_0 \rho_0)^{1/2}} \left(
\frac{(R^2 -r_{min}^2)^{1/2} (R^2 -\rho_0^2)^{1/2} }{R}
+\frac{\pi}{4} \left( \frac{\rho_0}{r_{min}} \right)^{1/2}r_{min}
\right)\,, \label{eq:tau-2}
\end{equation}
which becomes, assuming $R \gg r$ and $R \gg \rho_0$,
\begin{equation}
\tau \approx 2 \frac{R r_{min}}{\vF (3\pi\xi_0 \rho_0)^{1/2}}
=\frac{R}{\vF} \left( \frac{2
\Delta_0}{E-\Delta_0}\right)^{1/2}\approx
\frac{R}{\vF}\frac{2\Delta_0}{\ep}. \label{eq:tau-3}
\end{equation}
We conclude that the Andreev return time is longer if the
excitation's energy is lower.

Similarly, assuming $\rho_0/R \ll 1$ and $r_{min}/R \ll 1$,
the Andreev reflection angle is
\begin{equation}
\Delta \phi \approx \frac{\pi b}{ (3\pi\xi_0 \rho_0)^{1/2}}\,.
\label{eq:deltaphi-2}
\end{equation}

\begin{figure}
\centering \epsfig{figure=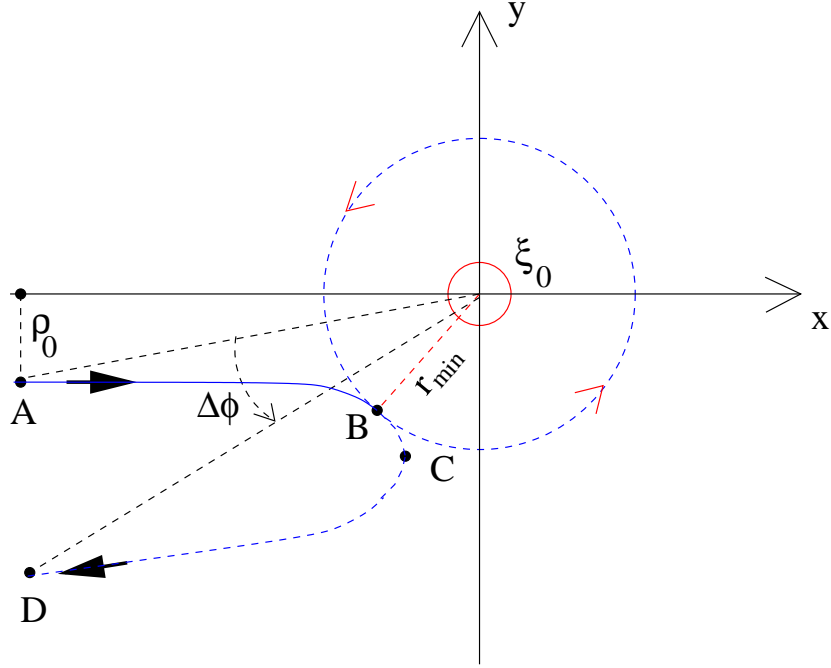,height=3.5in,angle=0}
\caption{(Color online) Schematic trajectory of the quasiparticle
which starts at position A, is Andreev--reflected by the vortex
(at the origin) at position B (where it becomes a quasihole), then
traces its way back with a small Andreev angle $\Delta \phi$ (not to scale).}
\label{fig:andreev}
\end{figure}

To apply these results we assume that the initial momentum of the
quasiparticle is directed along the $x$-axis, and that the angular
momentum $J=-py_0=\pF\rho_0$. From Eq.~(\ref{eq:ep}) it follows that
the momentum $p=\pF (1+2m^*\ep/\pF)^{1/2}$ and, in the ultra-low
temperature limit, $(p-\pF)/\pF\le 10^{-4}$. For $y_0$ we have
$y_0=\rho_0 (1+2m^*\ep/\pF)^{-1/2}\approx \rho_0$. In this case
$\rho_0$ becomes the impact parameter (Fig.~\ref{fig:andreev}), and
Eq.~(\ref{eq:deltaphi-2}) shows that quasiparticles with smaller
impact parameter (hence smaller angular momentum) are Andreev
reflected by smaller angles.

As it is seen from Eq.~(\ref{eq:rmin}), the Andreev radius depends
strongly on the initial energy of the excitation:
\begin{equation}
r_{min}=(3\pi\xi_0\rho_0)^{1/2}\frac{\Delta_0}{\ep}\,. \label{eq:rmin1}
\end{equation}

The same arguments apply to the critical value $\rho_{0c}$ defined
as a maximum value of $\rho_0$ which causes the Andreev reflection
of quasiparticles with the given initial energy $\ep$. To calculate
$\rho_{0c}$, we assume that at the starting point of the trajectory
the quasiparticle has coordinates $(R,\,\phi_0)$, where
$\phi_0=\arcsin(y_0/R)\approx -\arcsin(\rho_{0c}/R)$; the
coordinates of the Andreev reflection point in this case should be
$(r_{min},\,-\pi/2)$. Thus the difference between the reflection
angle and the starting angle is
\begin{equation}
\Delta\phi=-\frac{\pi}{2}+\arcsin\left(\frac{\rho_{0c}}{R}\right).
\label{eq:angdiff1}
\end{equation}
This difference can also be calculated from Eq.~(\ref{eq:dphi})
where the second term (of the order of $\hbar/\pF$) in the integrand
can be neglected. We obtain
\begin{equation}
\Delta{\phi}=-\arcsin\left(\frac{\rho_{0c}}{r_{min}}\right)
+\arcsin\left(\frac{\rho_{0c}}{R}\right)\,. \label{eq:angdiff2}
\end{equation}
By comparing Eqs.~(\ref{eq:angdiff1}) and (\ref{eq:angdiff2}) we
find
\begin{equation}
\rho_{0c} \approx 3\pi\xi_0 \left( \frac{\Delta_0}{\ep}\right)^2\,.
\label{eq:rho0c}
\end{equation}

In the typical low temperatures experiments $k_B T/\Delta_0 \approx
0.1$, and, for quasiparticles with initial energy $\ep\approx k_B
T$, we find $r_{min} \approx 10 (3\pi\xi_0 \rho_0)^{1/2}$ and
$\rho_{0c} \sim 10^3\xi_0$, while the same quantities for the
quasiparticles with $\ep\approx (\Delta_0 k_B T)^{1/2}$ are
$r_{min}\sim 3(3\pi\xi_0\rho_0)^{1/2}$ and $\rho_{0c} \sim 10^2\xi_0$.

\section{Heat transport through the velocity field of a vortex}

\renewcommand{\theequation}{\arabic{section}.\arabic{equation}}

In the experimental studies of superfluid turbulence in $^3$He-B at
the ultra-low temperatures the vortex tangle is studied by detecting
the fraction of quasiparticles which are Andreev reflected by the
vortices and measuring the heat which is transported by the
quasiparticles. Using the results of previous Sections, it is
straightforward to calculate the fraction of energy (or heat)
transmitted across the velocity field of a vortex. Once we know this
fraction, we shall generalize it to a system of many vortices.

In Section IV it was explained that the quasiparticles characterized
by the particular impact parameter $\rho_0$ are Andreev reflected by
a vortex if their energies satisfy the condition $\Delta_0\leq
E\leq\Delta_0 (1+3\pi\xi_0/2\rho_0)$. If this condition is not
satisfied, the quasiparticles pass freely across the vortex velocity
field. If the system which consists of the vortex and quasiparticles
is in thermal equilibrium, there is no preferred direction around
the vortex. Incident and transmitted fluxes at one side of the
vortex are canceled by the fluxes in the opposite direction, and no
net flow of energy exists when the temperature everywhere around the
vortex has the same value.

A net flux of quasiparticles and of energy results only if there is
some small temperature difference, $\delta T\ll T$ between the two
sides. If this is the case, the heat carried by the incident
quasiparticles is described by the expression:
\begin{equation}
\delta Q_{inc}=\int_{\Delta_0}^{\infty}N(E)v_g(E)E\frac{\partial
f(E)}{\partial T}\, \delta T\, dE, \label{eq:dQinc}
\end{equation}
where
\begin{equation}
N(E)=N_F\frac{E}{(E^2-\Delta^2)^{1/2}}\,. \label{eq:density}
\end{equation}
Here
\begin{equation}
N_F=\frac{mp_F}{\pi^2\hbar^3}
\end{equation}
is the density of states at the Fermi energy with corresponding
Fermi momentum $p_F$. The group velocity of a Bogolubov
quasiparticle $v_g$ is given by the expression:
\begin{equation}
v_g=\frac{\ep}{E}\vF=\frac{(E^2-\Delta^2)^{1/2}}{E}\vF \,, \label{eq:grVel}
\end{equation}
and $f(E)$ is the Fermi distribution function, which, at ultra--low
temperatures, is transformed into the Boltzman distribution, and
describes the mean occupation number of a state with energy $E$:
\begin{equation}
f(E)=e^{-\frac{E}{k_B T}}.
\label{eq:distrFunc}
\end{equation}

The thermal flux of quasiparticles incident on the vortex velocity
field per unit length per unit time is obtained with the help of
Eqs.~(\ref{eq:density}), (\ref{eq:grVel}) and (\ref{eq:distrFunc});
one has
\begin{equation}
\delta Q_{inc}=N_F v_F\frac{\delta T}{k_B T^2}
\int_{\Delta}^{\infty}E^2 e^{-\frac{E}{k_B T}}\,dE\approx N_F v_F
\Delta_0^2\frac{\delta T}{T}e^{-\frac{\Delta_0}{k_B T}}.
\label{eq:dincFlux}
\end{equation}
If there is a plane current of quasiparticles with transverse cross section
$R_0$, then the total heat current incident on the vortex per unit time will be:
\begin{equation}
Q_{inc}=2R_0\delta Q_{inc}=
2R_0 N_F v_F \Delta_0^2\frac{\delta T}{T}e^{-\frac{\Delta_0}{k_B T}}.
\label{eq:totherminc}
\end{equation}

We assume that, in the $(x,\,y)$-plane orthogonal to the straight
vortex line, the polarity of the vortex located at $(0,\,0)$ is
positive and consider quasiparticles incoming in the positive
$x-$direction. As discussed earlier, in this case the upper
half-plane will be absolutely transparent for quasiparticles so that
the heat transferred by quasiparticles through this half-plane will
meet no resistance. The lower half-plane of vortex flow field will
reflect a fraction of quasiparticles and induce some thermal
resistance. A quasiparticle with the impact parameter $\rho_0$ is
transmitted through the vortex velocity field if it carries the
energy $E>\Delta_0(1+\frac{3}{2}\pi\xi_0/\rho_0)$, in which case the
heat which is transferred per unit time by such a quasiparticle can
be calculated as
\begin{equation}
\delta Q(\rho)=\int_{\Delta_0(1+\frac{3\pi\xi_0}{2\rho_0})}^{\infty}
N(E)v_g(E)E\frac{\partial f(E)}{\partial T}\,\delta T\, dE\simeq
Q_{inc}\frac{1}{2R_0}\biggl(1+\frac{3\pi\xi_0}{\rho_0}\biggr)
e^{-\frac{\Delta_0}{k_B T}\frac{3\pi\xi_0}{2\rho_0}}.
\label{eq:transrho}
\end{equation}
Notice that estimating the ratio $\xi_0/\rho_0$ we kept only the
linear term.

The total amount of energy transferred through the vortex by
quasiparticles originated within the interval $-R_0\le y \le R_0$
is:
\begin{equation}
Q_{tr}=\frac{Q_{inc}}{2}\biggl[1+\frac{1}{R_0}\int_{0}^{R_0}
\left(1+\frac{3\pi\xi_0}{\rho_0}\right) e^{-\frac{\Delta_0}{k_B
T}\frac{3\pi\xi_0}{2\rho_0}}d\rho_0\biggr]\,. \label{eq:transheat}
\end{equation}
The integral in Eq.(\ref{eq:transheat}) can be estimated as
\begin{equation}
\approx R_0 e^{-\frac{\Delta_0}{k_B T}\frac{3\pi\xi_0}{2R_0}}.
\label{eq:intestim}
\end{equation}
Thus the fraction of heat which is transferred through the velocity
field of the vortex is
\begin{equation}
\delta f_{tr}=\frac{1}{2} \biggl(1+e^{-\frac{\Delta_0}{k_B
T}\frac{3\pi\xi_0}{2R_0}}\biggr). \label{eq:fractin}
\end{equation}

In experiments at ultra--low temperatures we have $\Delta_0/k_B
T\sim 10$, so that the cross-section of the thermal flux is
$R_0\sim 10\xi_0$ and approximately $52\%$ of the total heat is
transferred through the vortex. If the heat current has the
cross-section $\sim 10^2\xi_0$, the fraction of the transferred
heat is approximately $0.82$. Therefore the reflection of the heat
flux takes place only in the close vicinity of the vortex core.

\section{Andreev reflection in a vortex gas.}

\renewcommand{\theequation}{\arabic{section}.\arabic{equation}}

To apply our result to experiments, we consider for simplicity a system 
of random parallel-antiparallel vortices (i.e.
a system of vortex points in the $(x,\,y)$-plane; such a system is
known as the Onsager vortex gas). This vortex system is penetrated
by  a quasiparticle current created by a temperature difference
$\delta T$. It is convenient to introduce the effective radius $R_0$ of each
vortex as the half of the mean intervortex distance, i.e.
$R_0=\ell/2$. We divide the vortex configuration in parallel
layers of width $\ell$ each perpendicular to the quasiparticle
current. Clearly, the transmittability of each layer is equal to the
transmittability of a vortex within a region of radius $\ell/2$.
Thus the fraction of heat transmitted by each layer is
\begin{equation}
\delta f_{tr}=\frac{1}{2}\biggl(1+e^{-\frac{\Delta_0}{k_B T}
\frac{3\pi\xi_0}{\ell}} \biggr)\,. \label{eq:singlayer}
\end{equation}

If we assume now that vortices are well separated and that their
velocity fields do not overlap significantly, we obtain the
conditions
\begin{equation}
\xi_0\ll \ell, \quad\frac{\Delta_0}{k_B
T}\frac{3\pi\xi_0}{\ell}\ll 1. \label{wq:heat_conditions}
\end{equation}
Eq.~(\ref{eq:singlayer}) becomes
\begin{equation}
\delta f_{tr}\approx \frac{1}{2} \biggl(1+1-\frac{\Delta_0}{k_B
T}\frac{3\pi\xi_0}{\ell}\biggr)= 1-\frac{\Delta_0}{k_B
T}\frac{3\pi\xi_0}{2\ell}\approx e^{-\frac{\Delta_0}{k_B
T}\frac{3\pi\xi_0}{2\ell}}. \label{eq:singltrans}
\end{equation}

Driven by the temperature difference, the heat flux $Q_0$
reduces, after penetrating the first layer, to $Q_1=Q_0\,\delta
f_{tr}$; after penetrating the second layer, it becomes
$Q_2=Q_1\,\delta f_{tr}$. Hence, after penetrating the last $n$th
layer, we obtain $Q_n=Q_{n-1}\,\delta f_{tr}$. Thus we have
\begin{equation}
Q_n=Q_{n-1}\,\delta f_{tr}=...=Q_0\,\delta f_{tr}^n.
\label{eq:sequence}
\end{equation}
We conclude that the fraction of heat which is transferred through
the system of vortices is
\begin{equation}
f_{tr}= (\delta f_{tr})^n.
\label{eq:trsystem}
\end{equation}
If the total vorticity is confined within a region of size $S$,
the number of layers, $n$ can be estimated as a $n\approx S/\ell$.
From Eq.~(\ref{eq:trsystem}) we obtain:
\begin{equation}
f_{tr}=e^{-\frac{\Delta_0}{k_B T}\frac{3\pi\xi_0 S}{2\ell^2}}.
\label{eq:trsystem1}
\end{equation}
Finally we obtain the intervortex distance:
\begin{equation}
\ell=\biggl(-\frac{\Delta_0}{k_B T}\frac{3\pi\xi_0 S}{2\ln
f_{tr}}\biggr) ^{\frac{1}{2}}. \label{eq:interV}
\end{equation}
The quantities $S$ (the size of the vortex system) and $f_{tr}$ 
(the fraction of reflected quasiparticles) in Eq.~(\ref{eq:interV}) 
can be observed experimentally.  From the available description of 
one experiment\cite{Fisher}, the maximum transmitted fraction of
quasiparticle current is $f_{tr}\approx 0.75$ and the spatial
extension of the vorticity is $S\sim 2\cdot 10^{-1}\,\textrm{cm}$.
Since the zero temperature coherence length is $\xi_0\approx
0.8\times 10^{-5}\,\textrm{cm}$, we conclude that in the case
where $\Delta_0/k_B T\sim 10$ the average intervortex distance is
$\ell\sim 1.62\cdot 10^{-2}~\textrm{cm}$, which is in good
agreement with existing estimates\cite{Bradley1}.

\section{Conclusions}

In conclusion, starting from Hamilton's equations, we have calculated 
the trajectories of quasiparticles which move in the velocity field of 
a quantized vortex in $^3$He-B and determined the Andreev reflection point.
Generalizing the result to a disordered system of many vortices,
we have determined the precise location of turning point and showed 
how to recover the typical intervortex spacing in the turbulent $^3$He-B.
Our result is in good agreement with less rigorous estimates.

Future work will investigate Andreev reflection of quasiparticles
by a system of moving vortices. We shall also study how the Andreev reflection 
technique can be used to visualize vortex structures (e.g. coherent bundles of
vortices) and determine turbulent fluctuations and turbulence statistics.

\section{Acknowledgments}

This work was supported by EPSRC grant GR/T08876/01. NS also was supported by 
Georgian National Science Foundation 
grant GNSF/ST06/4-018.
We are also grateful to 
Professor S.N. Fisher for stimulating discussions and for reading the manuscript.

\vfill
\eject

\appendix
\section{}
\renewcommand{\theequation}{A.\arabic{equation}}

The Andreev return time, $\tau$, and the Andreev reflection angle,
$\Delta \phi$, are defined by formulae~(\ref{eq:tau-1}) and
(\ref{eq:deltaphi-1}), where $0 < \rho_0 < r_{min} < r < R$. To
evaluate these formulae we use the following integrals:
\begin{equation}
I_1=\int_{r_{min}}^R \frac{r^2\, dr}{ (r^2-r_{min}^2)^{1/2}
(r^2-\rho_0^2)^{1/2}}= \frac{\sqrt{R^2-r_{min}^2}}{\sqrt{R^2
-\rho_0^2}}
 + r_{min} G,
\label{eq:I1}
\end{equation}
where
\begin{equation}
G=
 K\left(\frac{\rho_0 }{r_{min}}\right)
-F\left(\arcsin{\frac{r_{min}}{R}},\,\frac{\rho_0
}{r_{min}}\right)
-E\left(\frac{\pi}{2},\,\frac{\rho_0}{r_{min}}\right)
+E\left(\arcsin{\frac{r_{min}}{R}},\,\frac{\rho_0}{r_{min}}\right),
\label{eq:G}
\end{equation}
and
\begin{equation}
I_2=\int_{r_{min}}^R \frac{dr}{ (r^2-r_{min}^2)^{1/2}
(r^2-\rho_0^2)^{1/2}}= \frac{1 }{r_{min} } \left[
K\left(\frac{\rho_0 }{r_{min}}\right)
-F\left(\arcsin{\frac{r_{min} }{R
}},\,\frac{\rho_0}{r_{min}}\right) \right]\,, \label{eq:I2}
\end{equation}
where $K$, $F$ and $E$ are elliptic integrals, defined as
\begin{equation}
F(k,\,\theta)=\int_0^{\theta}\frac{d\phi}{\sqrt{1-k^2
\sin^2{(\phi)} }}\,, \label{eq:elliptic-F}
\end{equation}
\begin{equation}
K(k)=F\left(\frac{\pi }{2 },\,k\right), \label{eq:elliptic-K}
\end{equation}
\begin{equation}
E(k,\,\theta)=\int_0^{\theta} \sqrt{1-k^2
\sin^2{(\theta)}}\,d\phi, \label{eq:elliptic-E}
\end{equation}
with $\theta=\arcsin{(r_{min}/R)}$ and $k=\rho_0/R$.

For $k^2 <1$ the elliptic integrals~(\ref{eq:elliptic-F}),
(\ref{eq:elliptic-K}) and (\ref{eq:elliptic-E}) are represented by
the series
\begin{equation}
F(k,\,\theta)=\frac{2 \theta}{\pi}K(k)
-\sin{\theta}\cos{\theta}\frac{k^2 }{4 }+ ...\,,
\end{equation}
\begin{equation}
K(k)=\frac{\pi}{2}+\frac{\pi^2}{8}k^2 + ...\,,
\end{equation}
\begin{equation}
E(k,\,\theta)=\frac{2 \theta}{\pi}E(k) +\sin{\theta}
\cos{\theta}\frac{k^2}{4}+...\,,
\end{equation}
\begin{equation}
E\left(k,\,\frac{\pi}{2}\right)=E(k)=\frac{\pi}{2}-\frac{\pi}{8}k^2+...\,,
\end{equation}
using which we obtain
\begin{equation}
\tau \approx \frac{2 r_{min}} {\vF (3\pi\xi_0 \rho_0)^{1/2}}
\left[ \frac{ \sqrt{R^2-r_{min}^2} \sqrt{R^2-\rho_0^2} }{R}
+\frac{\pi}{4} \left( \frac{\rho_0}{r_{min}}  \right)^2 r_{min}
\right]\,.
\end{equation}

Assuming $R \gg r_{min}$, $R \gg \rho_0$ and $\rho_0<r_{min}$
we have
\begin{equation}
\tau \approx \frac{2 R}{\vF}\frac{r_{min}}{(3\pi\xi_0
\rho_0)^{1/2}}\,.
\end{equation}

Similarly,
\begin{equation}
\Delta \phi \approx \pi \frac{b}{(3\pi\xi_0 \rho_0)^{1/2}  } \ll 1.
\end{equation}

\vfill
\eject

\clearpage

\end{document}